\documentclass[acmsmall]{acmart}
\AtBeginDocument{%
  }

\setcopyright{rightsretained}
\acmJournal{JACM}
\acmYear{2025} \acmVolume{1} \acmNumber{1} \acmArticle{1} \acmMonth{1} \acmPrice{}\acmDOI{10.1145/3765705}

\begin{document}

%%
%% The "title" command has an optional parameter,
%% allowing the author to define a "short title" to be used in page headers.
\title{From Passive to Participatory: 
How Liberating Structures Can Revolutionize Our Conferences}

\author{Daniel Russo}
\authornote{Corresponding author.}
\email{daniel.russo@cs.aau.dk}
\orcid{0000-0001-7253-101X}
\affiliation{%
  %\institution{Aalborg University}
  \institution{Department of Computer Science, Aalborg University}
  \streetaddress{A.C. Meyers Vaenge 15, 2450}
  \city{Copenhagen}
  \country{Denmark}}
  
\author{Margaret-Anne Storey}
\email{mstorey@uvic.ca}
\orcid{0000-0003-2278-2536}
\affiliation{%
  \institution{Department of Computer Science, University of Victoria}
  \streetaddress{P. O. Box 3055, STN CSC}
  \city{Victoria BC}
  \country{Canada}
}

%%
%% By default, the full list of authors will be used in the page
%% headers. Often, this list is too long, and will overlap
%% other information printed in the page headers. This command allows
%% the author to define a more concise list
%% of authors' names for this purpose.
\renewcommand{\shortauthors}{Russo and Storey}

%%
%% This command processes the author and affiliation and title
%% information and builds the first part of the formatted document.
\begin{abstract}
Our conferences face a growing crisis: an overwhelming flood of submissions, increased reviewing burdens, and diminished opportunities for meaningful engagement. With AI making paper generation easier than ever, we must ask whether the current model fosters real innovation or simply incentivizes more publications.
This article advocates for a shift from passive paper presentations to interactive, participatory formats. We propose \textit{Liberating Structures}—facilitation techniques that promote collaboration and deeper intellectual exchange. By restructuring conferences into two tracks—one for generating new ideas and another for discussing established work—we can prioritize quality over quantity and reinvigorate academic gatherings. Embracing this change will ensure conferences remain spaces for real insight, creativity, and impactful collaboration in the AI era.
\end{abstract}

\maketitle

\section*{A Time for Change}

Academic conferences have been at the heart of emerging research in software engineering for over 45 years. But our reliance on conferences to share and spark new research is starting to show some cracks. With a flood of submissions and new tracks in recent years, staying aware of relevant research and reviewing submissions has become a significant burden. And let us be honest: The rise of generative AI tools makes the task of writing more papers faster than ever.  

Undeniably, we are at a crossroads we cannot ignore, and we must ask ourselves, are ``\textit{more papers really better}''? As we spend more and more time writing, revising, submitting, and reviewing papers, is our community's research becoming more or less relevant to society's challanges, and are we really pushing the boundaries of our fields? 
With this increase in papers and tracks, we should also ask ourselves how our experience attending conferences has changed. 
As we struggle to attend and passively listen to even a few paper presentations from the many parallel tracks, are we really learning effectively about others’ work? Are we engaging in deep conversations and debates that spark new ideas? Are we making meaningful connections and forming new collaborations?   

\textbf{As AI reshapes our profession, fostering meaningful human engagement is more essential than ever}. With the increasing scale of conferences, creating participatory spaces for genuine intellectual exchange becomes crucial to supporting deep engagement with emerging ideas. The surge in submissions, however, risks overwhelming innovation with a high volume of contributions that may, while numerous, lack significant novelty or impact. 
After more than 45 years and the disruptive ways of new technology, is it time to rethink our conference model? It may be time for a radical shift in organizing our conference events. By adopting more deliberate quality over quantity-focused approaches, we can aim to reinvigorate conferences and enhance what we gain intellectually from them. We can transform our academic gatherings into spaces prioritizing thoughtful, deep engagement over superficial busyness and quantitative publishing metrics. It is not about doing more; it’s about doing what matters most—and doing it well. 

Here, we propose Liberating Structures (LS) as a concrete way to achieve a shift from passive conference presentations to more conversation and thinking. The Liberating Structure set of facilitation methods~\cite{torbert1978educating,Lipmanowicz2015} makes it easier for attendees to collaborate, share ideas, and generate innovative solutions by breaking down hierarchical barriers and fostering open, creative dialogues. By embracing LS, we can transform conferences into spaces that inspire and generate impactful community-driven research that supports disruptive thinking and shapes collaboration and innovation~\cite{Storey2024}. We propose it is time to embrace a model that values quality over quantity and prioritizes depth over haste, ensuring that our contributions to the field of computing are genuinely significant. 

\section*{The Unsustainable Problem with Our Conferences}

Today’s software engineering researchers are easy to recognize by their long hours at work and their harried expression of trying to keep up with the need to publish more and more papers and simultaneously review dozens of papers. It is not unusual to overhear conference researchers mentioning that they are retreating to their hotel rooms to refine or review papers for the next conference or journal. When they do attend presentations, they only half listen while on their laptops because of the stress they feel about the work they are not doing. When they do their own research back home, they find it challenging to find blocks of time to focus and review the literature relevant to their work in the face of daunting reviewing commitments. This cycle of an ever-increasing paper publication, an ``arms race'' of sorts, raises serious concerns about sustainability—human, social, and professional.

On a human level, researchers face burnout, mental fatigue, and declining well-being as personal boundaries blur under constant demands. On a social level, the culture of overwork can erode meaningful collaboration and mentorship, deterring young talent and weakening the community’s cohesion. On a professional level, the emphasis on quantity undermines deep inquiry and thoughtful reviewing, threatening the integrity and quality of the field's output. Regarding sustainability, the impact of travel on our environment and unfairness in terms of who can afford to attend cannot be ignored. Furthermore, many researchers are burning out with the demands of paper and travel, and the visibility of this culture turns many talented people away. 

As a concrete example, the International Conference on Software Engineering (ICSE), to many the leading software engineering venue, saw a dramatic rise in submissions over the past decade. In 2014, ICSE received 496 submissions, which increased almost linearly to 691 by 2022. However, the recent surge has been even more pronounced, with submissions jumping to 798 in 2023 and exceeding 1,000 in 2024. 
Some may speculate that this exponential growth is partly due to AI tools making it faster to generate and manage submissions (although using AI for reviewing is discouraged, adding an imbalance to the process). We have heard some colleagues mention they review hundreds of papers every year. This work is appreciated, but does it come at the cost of the thoughtful work this researcher could do that would lead to more meaningful work by the community as a whole?  

\section*{Transforming Conferences with Liberating Structures}

To address these challenges, we need innovative solutions that foster more engaging and interactive environments. One promising approach is the adoption of Liberating Structures. These simple yet powerful methods can fundamentally change how we organize and participate in conferences, moving away from the conventional, often passive formats to more dynamic and inclusive ones.

Liberating Structures (LS) are easy-to-learn interaction techniques designed to distribute control and invite contributions from all participants. Unlike traditional meeting formats that centralize control and limit participation, LS ensure that everyone’s voice is heard and valued. Each LS protocol consists of five elements: the structuring invitation, space arrangement, participation distribution, group configurations, and a sequence of steps. These elements guide the group through a process that fosters creativity, inclusion, and collaboration. For instance, consider the LS method ``Impromptu Networking.'' In a conference session using this approach, the facilitator poses a relevant question, such as ``What is the most significant challenge in integrating AI into software engineering?'' Participants pair up to discuss their thoughts for two minutes before switching to new partners. In just a few rounds, everyone has shared ideas with multiple people, quickly surfacing diverse perspectives and common themes.

The primary benefit of Liberating Structures is their ability to transform interactions in meetings, workshops, and conferences. By breaking down hierarchical barriers and encouraging equal participation, LS creates an environment where new ideas emerge, and meaningful connections can form. This leads to increased innovation, better problem-solving, and a stronger sense of community among participants. LS also makes meetings more engaging and productive by ensuring everyone can contribute and making the process more dynamic and interactive.

The software engineering community has already started experimenting with Liberating Structures to address the limitations of traditional conference formats. For example, a workshop at the Foundations of Software Engineering conference (FSE) focused on the ``2030 Roadmap for Software Engineering'' special issue of \textit{ACM Transactions on Software Engineering}. This workshop used LS to enhance the cross pollination of ideas from existing papers and to generate new collaborations for future research topics. In parallel, other communities such as NASA have found that structures like \textit{Critical Uncertainties}, \textit{Ecocycle Planning}, and \textit{Purpose to Practice} are particularly effective when the aim includes fostering innovation and interdisciplinary collaboration \cite{McCandless2024}. Together, these experiences demonstrate how LS can facilitate both immediate and long term collaboration across diverse domains.

Another example is the Copenhagen Symposium on Human-Centered Software Engineering AI (held in 2023 and again in 2024). The Symposium aimed to identify relevant research topics within the community and establish ongoing research workstreams. Using LS, participants committed to working remotely on chosen topics throughout the year, with plans to reconvene the following year to discuss their progress. This continuous cycle of collaboration and feedback, supported by the Sloan Foundation, exemplifies how LS can sustain engagement and innovation. For instance, through iterative processes, participants at the Symposium in 2023 co-developed the Copenhagen Manifesto~\cite{russo2024}. This manifesto focuses on the responsible integration of Generative AI and highlights how collective dialogue can produce shared visions that emphasize ethical, transparent, and human-centered approaches to technology. The Symposium in 2024 was even more effective, as many participants knew what to expect after using LS in 2023 and thus engaged more quickly. 

\section*{A Proposal for a New Model for Conferences}

Building on the success of such experiences, we can envision a new conference model that ensures they remain relevant and impactful in the future. This model includes two main tracks: the \emph{Creation Track} and the \emph{Sharing Track}. This approach leverages the power of Liberating Structures to promote a healthier, more engaged academic ecosystem.

In the \emph{Creation Track}, research topics emerge organically through LS techniques. Participants engage in dynamic, interactive sessions designed to spark new ideas and collaborations. Techniques like Impromptu Networking, 1-2-4-All, and Troika Consulting~\cite{LiberatingStructures} allow attendees to brainstorm, refine, and develop research topics in real-time. This track encourages participants to explore their shared interests deeply and commit to collaborative research projects. By fostering an environment of creativity and inclusion, the Creation Track ensures that the ideas generated are diverse, innovative, and aligned with the broader community's interests.

The \emph{Sharing Track} focuses on presenting and discussing mature, impactful ideas already published in journals or proceedings. It is designed to share fully developed research findings, best practices, and significant advancements with the broader community. Sessions in the Sharing Track may include LS-enhanced formats like Fishbowl Conversations and Conversation Café, encouraging active participation and deeper engagement from the audience. By presenting mature ideas in a structured yet interactive format, the Sharing Track ensures that conference content is both relevant and high-quality. This track allows researchers to showcase their work, receive valuable feedback, and inspire further innovations within the community, providing a platform for celebrating successes and learning from each other’s experiences. 

We recognize that small, deliberate steps will be necessary to ensure a smooth transition from the current conference model to a more interactive and impactful one. Introducing Liberating Structures (LS) in the form of pilot sessions within traditional conferences can be a first step, allowing participants to experience formats like Fishbowl Conversations and Conversation Café alongside standard presentations. Over time, these interactive methods can foster deeper engagement and collaboration, encouraging a cultural shift toward quality over quantity. This gradual approach can help address concerns around current incentive structures, where some institutions reward researchers simply for publishing in prestigious venues like ICSE. We can promote long-term innovation and meaningful progress by gradually realigning conference success metrics to emphasize impactful, community-driven contributions. This shift, facilitated by the Creation and Sharing Tracks, will transform conferences into dynamic hubs where real breakthroughs are celebrated.

\section*{Concluding Thoughts}

The time has come to reimagine our approach to academic conferences, ensuring they remain meaningful, impactful, and genuinely conducive to knowledge advancement. \textbf{By adopting Liberating Structures, we can cultivate more inclusive, productive, and engaging events—transforming conferences from passive gatherings into dynamic spaces for creativity, collaboration, and innovation}. As Cal Newport suggests, the path forward is not about doing more, but about doing what matters most with a focus on quality over quantity~\cite{Newport2023}. We recognize some challenges to changing what we do, but let us embrace this shift and foster a healthier academic ecosystem that values thoughtful research, meaningful connections, and impactful contributions.

\bibliographystyle{ACM-Reference-Format}
\bibliography{main}

\end{document}